\documentclass[conference]{IEEEtran}
\IEEEoverridecommandlockouts

\usepackage{cite}
\usepackage{amsmath,amssymb,amsfonts}
\usepackage{algorithmic}
\usepackage{graphicx}
\usepackage{textcomp}
\usepackage{xcolor}
\def\BibTeX{{\rm B\kern-.05em{\sc i\kern-.025em b}\kern-.08em
    T\kern-.1667em\lower.7ex\hbox{E}\kern-.125emX}}

\definecolor{myblue}{HTML}{0025AD}
\definecolor{mygreen}{HTML}{09AD00}

\usepackage{booktabs} 
\usepackage{url}
\usepackage{xspace}
\usepackage{amssymb}
\usepackage{fdsymbol}
\usepackage[font=normal]{caption}
\usepackage{subcaption}

\usepackage{float}
\usepackage{multirow,makecell}
\usepackage{pifont}
\usepackage{bbm}
\usepackage[inline]{enumitem}
\usepackage{adjustbox}

\usepackage{listings}
\usepackage{tikz}
\usepackage{pgfplots}
\pgfplotsset{compat=1.18}
\usetikzlibrary{tikzmark}
\usetikzmarklibrary{listings}

\usepackage{colortbl}
\definecolor{darkgreen}{rgb}{0.0, 0.3, 0.13}
\definecolor{darkred}{rgb}{0.2, 0.0, 0.13}

\usepackage{tcolorbox}
\tcbset{textmarker/.style={%
        parbox=false,boxrule=0mm,boxsep=0mm,arc=0mm,
        outer arc=0mm,left=3mm,right=3mm,top=3pt,bottom=3pt,
        toptitle=1mm,bottomtitle=1mm}
        }
        
\newtcolorbox{blueBox}{textmarker,
    colback=blue!10!white}

\usepackage{tabularx}

\usepackage{orcidlink}
\usepackage[utf8]{inputenc}
\usepackage{textgreek}
\DeclareUnicodeCharacter{2265}{\ensuremath{\geq}}

\newtheorem{takeaway}{Takeaway}

\usetikzlibrary{positioning, arrows.meta, shapes.geometric, calc}

\newcommand{\xmark}{\ding{55}}
\newcommand{\etal}{{\em et al.}\xspace}

\newcommand{\BfPara}[1]{{\noindent\bf #1.}\xspace}

\begin{document}

\title{\LARGE A Quasi-Experimental Developer Study of Security Training in LLM-Assisted Web Application Development\\

}
\author{Mohammed Kharma$^\heartsuit$, Ahmed Sabbah$^\heartsuit$, Radi Jarrar$^\heartsuit$, Samer Zain$^\heartsuit$, Mohammad Alkhanafseh$^\heartsuit$, David Mohaisen$^\vardiamondsuit$\\
$^\heartsuit$Birzeit University \hspace{10mm} $^\vardiamondsuit$University of Central Florida}
\maketitle

\begin{abstract}
This paper presents a controlled quasi-experimental developer study examining whether a layer-based security training package is associated with improved security quality in LLM-assisted implementation of an identity-centric Java Spring Boot backend. The study uses a mixed design with a within-subject pre-training versus post-training comparison and an exploratory between-subject expertise factor. Twelve developers completed matched runs under a common interface, fixed model configuration, counterbalanced task sets, and a shared starter project. Security outcomes were assessed via independent manual validation of submitted repositories by the first and second authors. The primary participant-level endpoint was a severity-weighted validated-weakness score. The post-training condition showed a significant paired reduction under an exact Wilcoxon signed-rank test ($p = 0.0059$). In aggregate, validated weaknesses decreased from 162 to 111 (31.5\%), the severity-weighted burden decreased from 432 to 267 (38.2\%), and critical findings decreased from 24 to 5 (79.2\%). The largest reductions were in authorization and object access (53.3\%) and in authentication, credential policy, and recovery weaknesses (44.7\%). Session and browser trust-boundary issues showed minimal change, while sensitive-data and cryptographic weaknesses showed only marginal improvement.

These results suggest that, under the tested conditions, post-training runs reduce validated security burden in LLM-assisted backend development without modifying the model. They do not support replacing secure defaults, static analysis, expert review, or operational hardening.

\end{abstract}


\section{Introduction}
\label{sec:introduction}

Large language models (LLMs) are increasingly used as coding assistants in software development~\cite{HouZLYWLLLGW23,HuangCCCPTHXZ24,JaffalAM25}. They help developers draft code, complete implementations, and explore alternatives. At the same time, recent evidence shows that security quality remains inconsistent, especially when security-relevant constraints are not explicitly specified and developers rely on default model behavior without structured guidance~\cite{PearceATDK22,PerrySKB23,KhouryABC23,TihanyiBFJC25}. This is particularly critical in web development, where weaknesses in authentication, authorization, credential management, file handling, and external system integration can have direct security impact~\cite{OWASPTop1024,CWEList4_6}.

A practical challenge is that secure code generation depends not only on the model, but also on how developers interact with it. Prompts often emphasize functional behavior while leaving assumptions and boundary conditions implicit~\cite{SandovalPNKGD23,ElgedawySDGGGJLLR24}. In such settings, the generated code may pass basic functionality tests and appear reasonable, yet still contain suboptimal implementation choices~\cite{KharmaCAD25}. Prior studies also show that developers do not consistently calibrate their trust in LLM-generated code and may rely on the assistant to varying degrees depending on task difficulty and context~\cite{PerrySKB23,AsareNA24,KlemmerHPLBPMRV24}. Consequently, ensuring secure LLM-assisted development is not only a matter of improving models, but also of shaping developer interaction with them. This paper studies a developer-centered bundled training package and examines whether developers produce more secure LLM-assisted code in the post-training condition. The training is grounded in the Mitigation-Aware Chain-of-Thought (MA-CoT) architectural-layer model~\cite{kharmaAM26}, which organizes security constraints by the layer where the primary mitigation operates. In this study, the model is used as a component of the training package and, secondarily, as a descriptive lens to interpret where validated vulnerabilities persist or decline across conditions.

To evaluate this question, we instantiate the study in a Java Spring Boot user-management backend, including security-sensitive functionality such as registration, login, password reset, multi-factor authentication, role-based administration, Active Directory integration, and document handling. We conduct a quasi-experimental controlled developer study with a within-subject pre-training versus post-training comparison and a between-subject expertise factor. All participants use the same blank Spring Boot starter project and interact with a fixed LLM configuration through a unified web interface. Security quality is evaluated via manual review by the authors using a predefined review guide aligned with the study scope~\cite{BrazBRCK22}.

\BfPara{Research Questions}
We evaluate the bundled training package through the following research questions:

\noindent\textbf{RQ1.} To what extent is the post-training condition associated with a lower number and severity of validated security weaknesses in LLM-assisted code generation?

\noindent{\em Rationale.} The primary objective is to assess whether structured, security-oriented prompting guidance is associated with improved security outcomes. Evaluating both counts and severity is necessary, as reductions in low-impact findings alone do not constitute meaningful improvement.

\noindent\textbf{RQ2.} How does developer expertise relate to the observed pre-training versus post-training difference in the security quality of LLM-generated code?

\noindent {\em Rationale.} The benefit of structured prompting support may vary with prior development experience and security knowledge. Given the limited sample, expertise is examined exploratorily to assess whether broad differences in improvement are observable across participant backgrounds.

These questions assess both the overall pre/post difference and its variation across levels of developer experience. This paper makes the following contributions: (1) \emph{Bundled training package.} We show that structured developer-side training reduces severity-weighted security weaknesses by 38.2\% and critical vulnerabilities by 79.2\% without modifying the underlying model.
(2) \emph{Controlled developer study with empirical results.} We design and conduct a quasi-experimental study with 12 developers and report paired pre-training versus post-training differences in validated security weaknesses, with expertise-stratified analyses treated as exploratory. (3) \emph{Theory-guided descriptive layer analysis.} We map each validated finding to a primary MA-CoT architectural layer and show that the pre/post difference is not uniform across layers, with the strongest reduction at the ecosystem library/framework layer. (4) \emph{Security-relevant task sets.} We construct two functional requirement sets grounded in realistic user-management workflows to elicit security-relevant implementation decisions without explicitly prescribing mitigation.



\BfPara{Organization} Section~\ref{sec:related_work} reviews the related work followed by our methodology in Section~\ref{sec:methodology}, the results and discussion in Section~\ref{sec:results}, threats to validity in Section~\ref{sec:threat_to_validity}, and concluding remarks and future work in Section~\ref{sec:conclusion}.

\section{Related Work}
\label{sec:related_work}

\BfPara{Security of LLM-Generated Code}
A growing body of empirical work shows that LLMs can generate insecure code in security-relevant tasks~\cite{PearceATDK22,KhouryABC23,FuLTLSYC25,MouDLZY25,TihanyiBFJC25,MousaviIMBA24,GongDTGYH25,hajipourHKFS23}. Recent surveys synthesize this evidence base~\cite{BasicG24,XuWLWZCYLW24}. Pearce~\etal evaluated GitHub Copilot on 89 vulnerability-oriented scenarios derived from the MITRE CWE Top 25 and found that about 40\% of 1,689 generated programs were vulnerable~\cite{PearceATDK22}. Khoury \etal evaluated ChatGPT-generated code on security-relevant programming problems and found that it often produced insecure programs, despite being able to recognize and explain vulnerabilities when explicitly prompted~\cite{KhouryABC23}. Sandoval \etal conducted a randomized user study with 58 participants on a C programming task and found that AI-assisted users produced critical security bugs at a rate no greater than 10\% above the control group~\cite{SandovalPNKGD23}. Extending beyond controlled prompts, Fu \etal analyzed 733 code snippets generated by GitHub Copilot, CodeWhisperer, and Codeium in GitHub projects and reported a high prevalence of security weaknesses~\cite{FuLTLSYC25}.
 
Tihanyi \etal compared nine LLMs on C code generation and found that at least 62.07\% of the 331{,}000 programs in FormAI-v2 were vulnerable, indicating consistently high vulnerability rates across models~\cite{TihanyiBFJC25}. Mousavi \etal examined Java security API usage by ChatGPT and found that around 70\% of generated code across 48 tasks contained API misuse, spanning 20 distinct types~\cite{MousaviIMBA24}, which is directly relevant to our Java-based setting. Mou \etal proposed CoV-Eval, a multi-task benchmark covering code completion, vulnerability repair, detection, and classification, and showed that LLMs continue to generate insecure code and struggle with repair and fine-grained vulnerability discrimination~\cite{MouDLZY25}. Gong \etal studied secure code generation and repair in Python and reported that more than 75\% of generated programs were vulnerable, while repair success ranged from 33.2\% to 59.6\%, with models performing worst on self-repair~\cite{GongDTGYH25}.  These studies motivate empirical work that examines both models and developer interaction patterns that shape the generated code.

\BfPara{Human Factors in LLM-Assisted Secure Coding}
Model-centric evaluations do not fully explain security outcomes in practice. Perry \etal conducted a randomized study with 47 participants and found that AI-assisted users produced less secure code and were more likely to overestimate its security~\cite{PerrySKB23}. In contrast, Asare \etal found that GitHub Copilot was associated with more secure solutions on harder tasks, with no effect on easier ones~\cite{AsareNA24}.  A Comparative evaluation of vulnerability-detection tasks also shows that LLM performance related to security remains uneven between model families, even when the task is explicitly framed around vulnerability reasoning~\cite{LinM25}. Klemmer \etal, through interviews and online discussions, showed that practitioners already use AI assistants for security-critical tasks and tend to verify outputs due to quality concerns~\cite{KlemmerHPLBPMRV24}. Elgedawy \etal further showed that security outcomes vary across models and usage contexts~\cite{ElgedawySDGGGJLLR24}. Braz and Bacchelli showed that security is rarely the primary focus during code review~\cite{BrazBRCK22}. Overall, these findings indicate that security outcomes depend not only on model capability, but also on task design, developer expertise, and interaction patterns.

\BfPara{Prompting and Intervention-Oriented Work}
Recent work shows that prompting strategy can influence the security of generated code. Tony \etal evaluate multiple prompting techniques across GPT-3, GPT-3.5, and GPT-4 and show reduced security weaknesses relative to baseline prompting~\cite{TonyFMDS25}. Bruni \etal similarly find that certain prompt-engineering strategies substantially reduce vulnerabilities, though effectiveness varies by model and task~\cite{BruniGGK25}. Nazzal \etal propose PromSec, an automated prompt-optimization framework that improves security while preserving functionality~\cite{NazzalKKP24}, and Zhao \etal report over 65\% vulnerability mitigation using a CWE-informed generation framework~\cite{ZhaoSHLGZL25}.

Other approaches incorporate external knowledge. Lin \etal use retrieval-augmented generation to inject vulnerability knowledge~\cite{LinWQCM25}, while Nong \etal show that chain-of-thought prompting improves vulnerability analysis and repair~\cite{NongACHCC24}. Le \etal demonstrate that effective vulnerability repair in JavaScript requires sufficient contextual information in prompts~\cite{LeAK24}. Cotroneo \etal introduce DeVAIC, a tool for evaluating vulnerabilities in AI-generated code~\cite{CotroneoLL25}.

Most prior work focuses on prompt design or evaluation tools, rather than whether developers can be trained to express security constraints during development. This distinction matters, as improvements under researcher-designed prompts may not transfer to typical developer behavior.

\BfPara{Security Benchmarks and Evaluation Frameworks}
Recent work has advanced evaluation methodologies for secure code generation. Siddiq and Santos introduced SecurityEval, a dataset of 130 Python samples spanning 75 CWE-mapped vulnerability types~\cite{SiddiqS22}. SALLM later proposed a systematic evaluation pipeline combining security-centric prompts with automated analysis~\cite{SiddiqSDM24}. Tony \etal presented LLMSecEval, a dataset of 150 natural-language prompts for assessing code security~\cite{tony2023llmseceval}. CWEval evaluates both functional correctness and security, showing that models can pass tests while still producing vulnerable code~\cite{PengCHYR25}. BaxBench uses exploit-based testing for backend code and shows that current models struggle to achieve both correctness and security~\cite{VeroMCRBJHV25}. While these benchmarks provide useful context, they remain model-centric and do not address how developer behavior can be shaped to improve outcomes.

\BfPara{Gap} Existing work demonstrates that prompting strategies and external knowledge can improve the security of generated code. However, these approaches assume that prompts are either engineered by researchers or optimized automatically. They do not address whether developers themselves can be trained to produce such prompts consistently in realistic development settings. This distinction is critical, as improvements observed under controlled conditions may not transfer to everyday practice. Our work addresses this gap by evaluating a structured training intervention that enables developers to express security constraints during LLM-assisted development.

\BfPara{Our Work}
Prior work largely treats secure code generation as a model or prompt-design problem, often relying on researcher-crafted prompts or automated prompt optimization. In contrast, this work focuses on the developer as the locus of control. We study whether developers can be trained to systematically express security constraints in a way that improves LLM outputs under realistic conditions. This represents a shift from model-centric alignment to developer-side alignment, where improvements arise from structured interaction rather than model modification.

\section{Methodology}
\label{sec:methodology}
The purpose of this study is to assess whether training developers in a layer-based security model improves the security and quality of code produced with LLM assistance. The core hypothesis is that developers generate safer code after receiving such training, which teaches them to include security constraints in prompts and reason about appropriate controls across software layers. We ground the study in a Java-based backend user management system to ensure realism and testable tasks, while focusing on generalizable interaction effects rather than framework-specific behavior.

\subsection{Study Design}
\begin{figure}
    \centering
\begin{tikzpicture}[
    scale=0.6,
    transform shape,
    node distance=0.8cm and 0.5cm,
    box/.style={
        draw, 
        rectangle, 
        align=center, 
        font=\sffamily\small, 
        line width=0.5pt,
        fill=mygreen!40,
        inner sep=5pt
    },
    graybox/.style={
        box, 
        fill=myblue!10, 
        draw=none
    },
    titlebox/.style={
        box, 
        fill=mygreen!10, 
        text width=14cm, 
        font=\sffamily\bfseries\small
    },
    arrow/.style={
        -{Stealth[scale=1.2]}, 
        thick
    }
]

\node[titlebox] (recruitment) {Participant Recruitment (N = 12 developers) - Between-Subject Factor: Expertise Level};

\node[graybox, below left=0.1cm and -3.5cm of recruitment, anchor=north] (junior) {Junior Developers ($\le$ 5 years experience; n = 6)};
\node[graybox, below right=0.1cm and -3.5cm of recruitment, anchor=north] (senior) {Senior Developers ($\ge$ 6 years experience; n = 6)};

\node[titlebox, below=1.0cm of recruitment] (consent) {Consent \& Background Survey + Environment Setup};

\node[titlebox, below=0.5cm of consent] (counterbalance) {Counterbalanced Task Assignment (within expertise group)\\ \textit{Schedule 1: Set A $\rightarrow$ Set B | Schedule 2: Set B $\rightarrow$ Set A}};

\node[draw, rectangle, inner sep=15pt, below=0.5cm of counterbalance, minimum width=14.5cm, minimum height=3.8cm] (container) {};

\node[graybox, below left=0.6cm and -2.5cm of counterbalance, anchor=north, text width=4.5cm] (pre1) {
    \textbf{Schedule 1 ($n$$=$$6$: 3 Jr + 3 Sr)}\\
    \textbf{Run 1: Pre-Training}\\
    \textbf{Task Set A}\\
    \textbf{Start: $S_0$ snapshot}
};

\node[graybox, below=0.2cm of pre1, text width=4.5cm] (pre2) {
    \textbf{Schedule 2 ($n$$=$$6$: 3 Jr + 3 Sr)}\\
    \textbf{Run 1: Pre-Training}\\
    \textbf{Task Set B}\\
    \textbf{Start: $S_0$ snapshot}
};

\node[graybox, rotate=-90, right=3.5cm of $(pre1.south)!0.5!(pre2.north)$, anchor=center, text width=3cm] (intervention) {Training Intervention};

\node[graybox, right=1.2cm of intervention, yshift=2.7cm, text width=3.5cm] (post1) {
    \textbf{Run 2: Post-Training}\\
    \textbf{Task Set A}\\
    \textbf{Start: $S_0$ snapshot}
};

\node[graybox, right=1.2cm of intervention, yshift=0.7cm, text width=3.5cm] (post2) {
    \textbf{Run 2: Post-Training}\\
    \textbf{Task Set A}\\
    \textbf{Start: $S_0$ snapshot}
};

\node[graybox, right=2.65cm of $(post1.south)!0.5!(post2.north)$, anchor=west, text width=2cm] (eval) {
    \textbf{Evaluation Metric}\\
    1) Security review\\
    2) Pre vs Post
};

\draw[arrow] (recruitment) -- (consent);
\draw[arrow] (consent) -- (counterbalance);
\draw[arrow] (counterbalance) -- (container.north);

\draw[arrow] (pre1.east) -- ++(0.2,0) |- (intervention.south);
\draw[arrow] (pre2.east) -- ++(0.2,0) |- (intervention.south);

\draw[arrow] (intervention.north) -- ++(0.2,0) |- (post1.west);
\draw[arrow] (intervention.north) -- ++(0.2,0) |- (post2.west);

\draw[arrow] (post1.east) -- ++(0.2,0) |- (eval.west);
\draw[arrow] (post2.east) -- ++(0.2,0) |- (eval.west);

\end{tikzpicture}
\caption{Mixed-design developer study with a within-subject training factor and a between-subject expertise factor. Each participant completed pre- and post-training runs on counterbalanced task sets, both starting from the same clean snapshot and evaluated with the same security review workflow.}    \label{fig:user_study_design}\vspace{-3mm}
\end{figure}
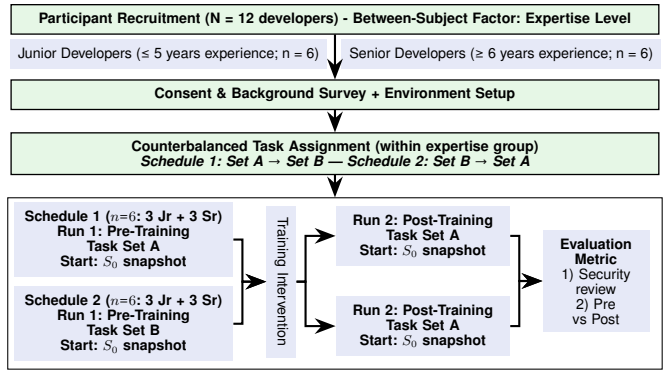

The study uses a controlled quasi-experimental mixed design with a within-subject factor and a between-subject factor, as shown in Fig.~\ref{fig:user_study_design}. The figure illustrates participant grouping, counterbalanced task assignment, the structure of pre- and post-training runs, and the evaluation procedure. The within-subject design increases statistical sensitivity by allowing each participant to serve as their own control, reducing variance due to individual differences in coding ability and security knowledge. This is particularly important given the modest sample size and strengthens the reliability of observed effects.

\subsubsection{Training-package condition}
Each participant completed two development runs. In the \emph{pre-training (baseline)} condition, participants used free-form prompting without study-specific guidance. In the \emph{post-training} condition, participants first received a standardized training package and then completed the tasks using the structured prompting method.

\subsubsection{Developer expertise}
To examine whether the pre/post difference varies by experience, participants were grouped into two levels: \emph{junior} developers (up to five years of professional experience) and \emph{senior} developers (six or more years).

\subsection{Layer-Based Training Content}
The post-training condition uses a bundled package with three components: (i) security-oriented instructional content, (ii) a layer-based mental model for reasoning about where mitigations operate, and (iii) a structured prompting scaffold. The model distinguishes: (1) \emph{Language-core (strict):} weaknesses from language semantics, e.g., integer overflow in Java arithmetic without bounds checks; (2) \emph{Standard runtime or library:} weaknesses from built-in APIs or defaults, e.g., unsafe XML parsing when \texttt{DocumentBuilderFactory} does not disable external entity resolution; (3) \emph{Ecosystem library or framework:} weaknesses from third-party frameworks, e.g., overly permissive Spring Security rules exposing protected endpoints; (4) \emph{Platform or OS API:} weaknesses from OS primitives, e.g., time-of-check to time-of-use races in file handling; (5) \emph{Toolchain or optimization:} weaknesses from compiler or runtime behavior, e.g., relying on \texttt{assert} for security checks when assertions are disabled in production; (6) \emph{Application security logic or policy:} weaknesses from app-level decisions, e.g., missing checks or weak policies.

Participants are trained to incorporate this model into structured prompts that specify: (i) functional intent, (ii) environment constraints (language and framework versions), (iii) boundary expectations (authentication, authorization, input validation, error handling), and (iv) explicit security assumptions aligned with the layer checklist.

The training package is evaluated as an integrated intervention rather than as isolated components. This reflects realistic deployment scenarios, where instructional material, mental models, and prompting scaffolds are used jointly. While this design does not isolate the contribution of individual components, it provides a more ecologically valid assessment of developer-side interventions.

\subsection{Counterbalancing and Use of Task Sets A and B}
Because the training package is not reversible, the study uses a fixed order with pre-training preceding post-training. To mitigate order effects, task-set assignment is counterbalanced. Under Schedule 1, Run 1 (\emph{pre-training}) uses task set A and Run 2 (\emph{post-training}) uses task set B; under Schedule 2, the order is reversed. To limit carry-over from implementation artifacts, each run starts from the same clean project snapshot ($S_0$) and is evaluated as an independent repository state.

\subsection{Participants and Sample Size}
We target 12 participants (6 junior, 6 senior), balancing resource constraints with sensitivity from the within-subject design, where each participant serves as their own control. We report effect sizes for paired outcomes and interpret results conservatively. All participants follow a standardized procedure, including a brief background questionnaire (development experience, Java/backend experience, LLM usage, prior security training) and a training-duration log capturing start and end times. This reduces procedural variation and ensures consistent metadata collection.


\BfPara{Ethics and Consent}
All participants provide informed written consent. The consent form explains the study purpose, data collection scope, anonymization procedures, and the right to withdraw at any time without consequence. No personally identifiable information is included in the analysis or reported results. All data is stored in an access-controlled environment and referenced only through anonymous participant IDs.

\subsection{Controlled Setup and Unified LLM Environment}
All sessions are conducted in person using pre-configured laptops. To reduce configuration variance, participants do not use personal tools or accounts. Instead, we provide a unified web interface that connects all participants to the same fixed model version, \texttt{gpt-5.2-2025-12-11}~\cite{GPT5OpenAI25}, under a common system configuration. This standardizes interaction and avoids differences due to model selection, prompt settings, or client-side features.

\subsection{Task-Sets}
The study uses two task sets, \emph{task-set A} and \emph{task-set B}, alternated across runs according to the counterbalancing schedule. In both conditions, participants implement features for an identity-centric user-management backend using a controlled Spring Boot starter project. The starter provides a common structure, configuration baseline, and package layout, but no feature implementation code, ensuring a shared starting point while preserving participant control over security-relevant decisions. Each run starts from a clean project snapshot and is evaluated independently. Participants receive requirements in software-engineering style using functional statements of the form ``The system shall \ldots'' to reflect realistic feature requests. Although framed functionally, tasks are designed to elicit security-relevant decisions suitable for expert review. Acceptance tests are available during development for progress checking and consistency, but serve only as a completion aid.

\BfPara{Task-Set Calibration}
Task-sets A and B were designed to be comparable in implementation effort and security relevance while covering distinct, realistic classes of backend weaknesses. Calibration considered the number of externally reachable endpoints, architectural layers exercised, amount of stateful logic, presence of security-sensitive operations (e.g., cryptography, parsing, query handling), and expected code volume and integration overhead. Both sets were implemented from the same clean snapshot ($S_0$) under identical toolchain, dependency, interface, and LLM settings. The sets were piloted and iteratively refined until completion effort and coverage were broadly comparable; pilot submissions were excluded from the analysis.

\subsection{Evaluation Methodology}
The first two authors conducted the final security evaluation. Both have prior experience in application and backend security and are familiar with Spring Boot. They used a predefined review guide and performed independent assessments. Each participant's final snapshot was manually reviewed. All 24 submissions were evaluated using a guide covering authentication and session management, authorization, input validation and sanitization, error handling and information exposure, credential and secret management, file and document handling, and external-system integration. For each confirmed weakness, reviewers recorded the CWE ID, severity, affected location, a brief description, and a mitigation rationale.

Disagreements were resolved through structured adjudication. Findings raised by only one reviewer were re-examined and confirmed or rejected using the guide. For agreed findings with differing severity, reviewers discussed exploitability, impact, and context until assigning a single severity; if unresolved, the more conservative rating was used.

\BfPara{Analysis Plan}
The primary analysis is at the participant level, with each participant contributing one submission per condition. Inferential claims are based on within-participant change. For participant $i$ in condition $c \in \{\mathrm{pre}, \mathrm{post}\}$, let $C_i^{(c)}$, $H_i^{(c)}$, $M_i^{(c)}$, and $L_i^{(c)}$ denote counts of Critical, High, Medium, and Low weaknesses. The main endpoint is the severity-weighted score
\begin{equation}
S_i^{(c)} = 4C_i^{(c)} + 3H_i^{(c)} + 2M_i^{(c)} + L_i^{(c)},
\label{eq:weighted_score}
\end{equation}
and the corroborative outcome is the total count
\begin{equation}
T_i^{(c)} = C_i^{(c)} + H_i^{(c)} + M_i^{(c)} + L_i^{(c)}.
\label{eq:total_score}
\end{equation}
Paired change scores are
\begin{equation}
\Delta S_i = S_i^{(\mathrm{pre})} - S_i^{(\mathrm{post})},
\label{eq:delta_weighted}
\end{equation}
and
\begin{equation}
\Delta T_i = T_i^{(\mathrm{pre})} - T_i^{(\mathrm{post})},
\label{eq:delta_total}
\end{equation}
so positive values indicate improvement. Secondary outcomes include severity-stratified counts and descriptive weakness-family distributions. Given the paired design and modest sample, within-participant comparisons use exact Wilcoxon signed-rank tests, with Holm adjustment across confirmatory tests. Expertise- and severity-stratified analyses are exploratory; between-group comparisons use descriptive summaries and Mann--Whitney tests on change scores. Acceptance tests and build checks are used only to confirm evaluable submissions and are not endpoints. As a design check, we also compare baseline burden and improvement across counterbalancing schedules descriptively.

To reduce subjective bias, reviewers conducted independent assessments using a predefined guide, followed by structured adjudication with conservative resolution rules.

\BfPara{Severity Weight Justification}
The 4/3/2/1 weighting scheme was selected as the primary aggregation rule because it preserves severity ordering while remaining interpretable and not sensitive to extreme cases. To assess robustness, we performed a sensitivity analysis using an exponential scheme (10/5/2/1), which emphasizes critical findings, and compared results with the unweighted total-count outcome in Eq.~\eqref{eq:total_score}. The paired Wilcoxon signed-rank test remained significant for the primary weighted score ($p = 0.0059$), the exponential scheme ($p = 0.0034$), and the unweighted outcome ($p = 0.0371$), indicating robustness to the aggregation rule.

\BfPara{Weakness-Family Aggregation}
Each CWE was mapped to a single researcher-defined weakness family. Families capture recurring security concerns in identity-centric backends while reducing fragmentation across low-frequency CWEs. This aggregation is used only for descriptive interpretation; all primary inference is based on participant-level validated counts and the severity-weighted endpoint. The full CWE-to-family mapping is provided in the appendix.

\BfPara{MA-CoT Layer Attribution}
As a secondary descriptive analysis, each confirmed finding was assigned to a primary MA-CoT layer based on the location of the main mitigation. Framework defaults and configuration issues were coded as \emph{Ecosystem library/framework}; application-specific authorization, logging, secret management, and policy decisions as \emph{Application security logic/policy}; transport or OS interface issues as \emph{Platform/OS API}; and misuse of built-in Java APIs as \emph{Standard runtime/library}.

\section{Results and Discussion}\label{sec:results}

Results are derived from the consolidated validated review report covering all 24 repositories (12 pre-training, 12 post-training). Repositories were mapped to participants and expertise groups to preserve the paired design. Unless noted, the severity-weighted score (Section~\ref{sec:methodology}) is the primary outcome, capturing both count and severity of validated weaknesses.

\BfPara{Counterbalancing Validation}
As a design check, baseline burden and weighted improvement were compared descriptively across the two counterbalancing schedules; no substantial imbalance was observed.

\begin{takeaway}
Counterbalancing checks provide no evidence that the observed improvement is attributable to task-set order rather than to the training package.
\end{takeaway}

\begin{figure}[t]
\centering
\begin{tikzpicture}
\begin{axis}[
    ybar,
    bar width=6pt,
    width=0.95\linewidth,
    height=3.5cm,
    ymin=0, ymax=80,
    ylabel={Severity score},
    xlabel={\small Participant},
    symbolic x coords={P01,P02,P03,P04,P05,P06,P07,P08,P09,P10,P11,P12},
    xtick=data,
    xlabel style={font=\small},
    ylabel style={font=\small},
    yticklabel style={font=\small},
    xticklabel style={rotate=45, anchor=east, font=\small},
    legend style={
        at={(0.98,0.98)},
        anchor=north east,
        legend columns=2,
        column sep=5pt
    },
    ymajorgrids=true,
    grid style=dashed,
    enlarge x limits=0.08
]

\addplot+[bar shift=-3pt, fill=myblue!40, draw=myblue] coordinates {
    (P01,23) (P02,47) (P03,36) (P04,49)
    (P05,43) (P06,39) (P07,37) (P08,25)
    (P09,44) (P10,28) (P11,27) (P12,34)
};

\addplot+[bar shift=3pt, fill=mygreen!40, draw=mygreen] coordinates {
    (P01,13) (P02,6) (P03,45) (P04,18)
    (P05,32) (P06,17) (P07,36) (P08,25)
    (P09,32) (P10,28) (P11,9) (P12,6)
};

\legend{Pre-training, Post-training}

\end{axis}
\end{tikzpicture}
    \caption{Participant-level severity-weighted scores before and after training. Nine of twelve participants show a reduction; one (P03) increases, and two (P08, P10) are unchanged. Bar heights correspond to Eq.~\eqref{eq:weighted_score}.}
    \label{fig:paired_bars}\vspace{-2mm}
\end{figure}
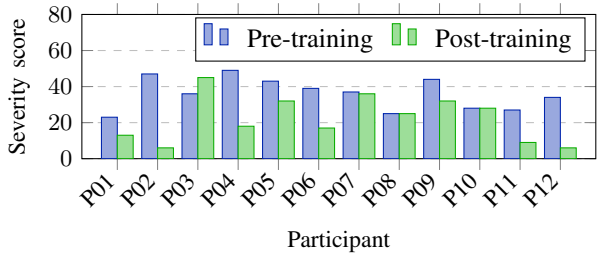

\BfPara{RQ1: Confirmatory Analysis of the Post-Training Association}
Table~\ref{tab:overall_results} reports aggregate totals. Validated findings decreased from 162 (pre) to 111 (post), a 31.5\% reduction. The severity-weighted score decreased from 432 to 267 (38.2\%). Critical findings declined from 24 to 5, and high-severity findings from 80 to 54. These trends favor the post-training condition; the inferential question is whether the improvement holds at the participant level. Table~\ref{tab:paired_outcomes} reports paired outcomes. The principal endpoint shows a significant reduction post-training, with a median improvement of 11.5 weighted points (exact Wilcoxon $p = 0.0059$, Holm-adjusted $p = 0.0117$) and a large effect size ($r_{rb}=0.93$). The mean weighted score decreased from 36.00 to 22.25. Fig.~\ref{fig:paired_bars} shows reductions across participants rather than a single outlier. Nine participants improved, one worsened, and two were unchanged.

The observed effect size (rrb = 0.93) indicates a near-uniform direction of improvement across participants, suggesting that the effect is not driven by a small number of outliers but reflects a consistent pattern.

\begin{table}[t]
\centering
\caption{Overall validated weakness outcomes.}
\label{tab:overall_results}
\scalebox{0.92}{
\begin{tabular}{lrrrrrr}
\toprule
Condition & Total & Weighted & Critical & High & Med. & Low \\
\midrule
Pre-training & 162 & 432 & 24 & 80 & 38 & 20 \\
Post-training & 111 & 267 & 5 & 54 & 33 & 19 \\
Change & -31.5\% & -38.2\% & -79.2\% & -32.5\% & -13.2\% & -5.0\% \\
\bottomrule
\end{tabular}
}
\end{table}

The unweighted result shows the same direction as the principal weighted endpoint. The median paired reduction in total findings was 4.0 (exact Wilcoxon $p = 0.0371$, Holm-adjusted $p = 0.0371$), with a large effect size ($r_{rb}=0.76$). We treat the total-count result as supporting evidence, not a replacement for the weighted analysis. Severity-stratified counts follow a similar pattern: critical findings decline with a large effect, and high-severity findings trend downward, while medium- and low-severity findings change little. This indicates that the post-training condition reduces both overall weaknesses and the burden of severe failures. The appendix reports participant-level severity composition and provides a full audit trail for the paired analysis, showing that improvement is distributed across participants rather than driven by a single outlier.

\begin{table}[t]
\centering
\caption{Participant-level paired security outcomes before and after training. The table reports pre-training and post-training means, median paired change ($\Delta = \mathrm{pre} - \mathrm{post}$), Wilcoxon signed-rank $p$-values, Holm-adjusted $p$-values for confirmatory outcomes, and matched-pairs rank-biserial effect sizes. Positive $\Delta$ indicates improvement.}
\label{tab:paired_outcomes}
\scalebox{0.84}{
\begin{tabular}{lrrrrrr}
\toprule
Outcome & Pre mean & Post mean & Median $\Delta$ & $p$ & Holm $p$ & $r_{rb}$ \\
\midrule
Weighted score & 36.00 & 22.25 & 11.5 & 0.0059 & 0.0117 & 0.93 \\
Total findings & 13.50 & 9.25 & 4.0 & 0.0371 & 0.0371 & 0.76 \\
Critical & 2.00 & 0.42 & 1.5 & 0.0117 & --- & 0.91 \\
High & 6.67 & 4.50 & 2.5 & 0.0522 & --- & 0.65 \\
Medium & 3.17 & 2.75 & 0.0 & 0.6523 & --- & 0.22 \\
Low & 1.67 & 1.58 & 1.0 & 0.4961 & --- & 0.29 \\
\bottomrule
\end{tabular}
}\vspace{-3mm}
\end{table}

While practice effects cannot be fully excluded due to the fixed pre-to-post sequence, several factors suggest they are unlikely to fully explain the observed improvements: (i) counterbalanced task sets, (ii) independent resets to a clean project state, and (iii) the magnitude and consistency of participant-level improvements across severity-weighted outcomes.

\begin{takeaway}
The post-training condition is associated with a substantial reduction in validated security burden, with strongest evidence at the participant level on the severity-weighted endpoint ($p=0.0059$, $r_{rb}=0.93$).
\end{takeaway}

\BfPara{RQ2: Exploratory Analysis of Developer Expertise}
Table~\ref{tab:expertise_descriptive} provides a descriptive view by experience level. Junior participants show a larger median weighted improvement, while senior participants also improve and reduce aggregate critical findings. These patterns are descriptive and do not establish an expertise-by-package effect.

\begin{table}[t]
\centering
\caption{Descriptive results by expertise. Columns report the mean number of validated findings (pre to post), the median participant-level weighted improvement $\Delta S_i$ (Eq.~\eqref{eq:delta_weighted}), and the aggregate number of critical findings (pre to post).}
\label{tab:expertise_descriptive}
\scalebox{0.78}{
\begin{tabular}{lccc}
\toprule
Group & Mean findings (Pre $\rightarrow$ Post) & Median $\Delta$ weighted & Critical total (Pre $\rightarrow$ Post) \\
\midrule
Junior & 14.33 $\rightarrow$ 8.17 & 20.0 & 13 $\rightarrow$ 4 \\
Senior & 12.67 $\rightarrow$ 10.33 & 5.0 & 11 $\rightarrow$ 1 \\
\bottomrule
\end{tabular}
}\vspace{-3mm}
\end{table}

Between-group differences should be interpreted cautiously. All exact Mann--Whitney tests on participant-level change scores are non-significant ($p \geq 0.1797$). The appropriate conclusion is limited: both groups improve, but the sample is too small to support an expertise-by-package interaction.

\begin{takeaway}
Both groups improve descriptively, but the sample is too small to support an expertise-by-package claim.
\end{takeaway}

\BfPara{Participant-Level Exception (P03)}
Participant P03 was the only case with an increased severity-weighted score after training (36 to 45). This reflects a shift in weakness composition rather than a uniform decline: findings increased from 15 to 18, with a broader mix of access-control, logging-exposure, and configuration issues. P03 is retained to preserve the paired design and reflect the full participant-level distribution.

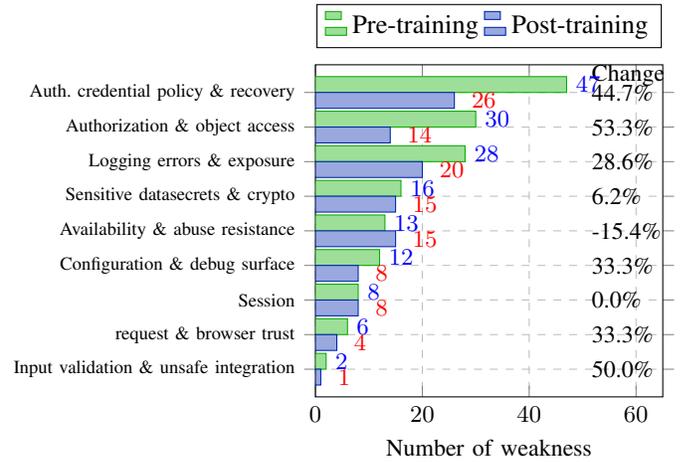
\begin{figure}[t]
\centering
\begin{tikzpicture}
\begin{axis}[
    xbar,
    width=0.7\linewidth,
    height=6cm,
    xmin=0, xmax=65,
    xlabel={Number of weakness},
    y dir=reverse,
    ytick={0,1,2,3,4,5,6,7,8},
    yticklabels={
        Auth. credential policy \& recovery,
        Authorization \& object access,
        Logging errors \& exposure,
        Sensitive datasecrets \& crypto,
        Availability \& abuse resistance,
        Configuration \& debug surface,
        Session, request \& browser trust,
        Input validation \& unsafe integration,
        File handling
    },
    yticklabel style={align=right, font=\scriptsize},
    xlabel style={font=\small},
    xticklabel style={font=\small},
    legend style={
        at={(0.5,1.05)},
        anchor=south,
        legend columns=2
    },
    grid=both,
    grid style=dashed,
]

\addplot+[
    xbar,
    bar width=6pt,
    bar shift=3pt,
    fill=mygreen!40,
    draw=mygreen,
    nodes near coords,
    every node near coord/.append style={
        font=\small,
        anchor=west
    }
] coordinates {
    (47,0) (30,1) (28,2) (16,3) (13,4) (12,5) (8,6) (6,7) (2,8)
};

\addplot+[
    xbar,
    bar width=6pt,
    bar shift=-3pt,
    fill=myblue!40,
    draw=myblue,
    nodes near coords,
    every node near coord/.append style={
        font=\small,
        anchor=west,
        xshift=3pt
    }
] coordinates {
    (26,0) (14,1) (20,2) (15,3) (15,4) (8,5) (8,6) (4,7) (1,8)
};

\legend{Pre-training, Post-training}

\node[anchor=west, font=\small] at (axis cs:50,0) {44.7\%};
\node[anchor=west, font=\small] at (axis cs:50,1) {53.3\%};
\node[anchor=west, font=\small] at (axis cs:50,2) {28.6\%};
\node[anchor=west, font=\small] at (axis cs:50,3) {6.2\%};
\node[anchor=west, font=\small] at (axis cs:50,4) {-15.4\%};
\node[anchor=west, font=\small] at (axis cs:50,5) {33.3\%};
\node[anchor=west, font=\small] at (axis cs:50,6) {0.0\%};
\node[anchor=west, font=\small] at (axis cs:50,7) {33.3\%};
\node[anchor=west, font=\small] at (axis cs:50,8) {50.0\%};

\node[anchor=west, font=\small] at (axis cs:50,-0.5) {Change};

\end{axis}
\end{tikzpicture}
    \caption{Changes in researcher-defined weakness families from pre- to post-training. For each family, the figure shows total counts before and after training, with percentage change relative to the pre-training count.}
    \label{fig:theme_shifts}
\end{figure}

\BfPara{Domain-Specific Weakness Patterns}
For domain-level interpretation, each CWE was mapped to a single researcher-defined weakness family and analyzed descriptively. Fig.~\ref{fig:theme_shifts} highlights the largest pre/post differences. The largest proportional reduction appears in authorization and object access (30 to 14, 53.3\%), followed by authentication, credential policy, and recovery (47 to 26, 44.7\%). These core failure modes in identity-centric systems were emphasized in training (boundary assumptions, ownership checks, route protection, credential handling, recovery flows). Logging, error, and information-exposure weaknesses decreased (28 to 20, 28.6\%), as did configuration and debug-surface issues (12 to 8, 33.3\%).

In contrast, session, request, and browser trust-boundary weaknesses show no change (8 to 8), while sensitive-data, secret-handling, and cryptography issues improve marginally (16 to 15, 6.2\%). Availability and abuse-resistance do not improve (13 to 15, $-15.4\%$). Overall, improvements are larger for access control, boundary control, and credential policy than for operational controls, session boundaries, and areas dependent on secure defaults or framework hardening.

We notice that improvements are concentrated in areas where security constraints can be explicitly expressed through prompting (e.g., authorization and credential handling), whereas limited gains in cryptographic and operational controls suggest reliance on framework defaults and deeper expertise beyond prompt-level guidance.

\begin{takeaway}
Largest gains occur in authorization and authentication/recovery, while operational hardening and browser trust-boundary weaknesses remain persistent.
\end{takeaway}

\BfPara{MA-CoT Layer Attribution}
Each of the 273 confirmed findings is assigned to a primary MA-CoT layer based on the location of the main mitigation. This mapping is used to interpret where pre/post differences concentrate.

Table~\ref{tab:layer_attribution} shows the largest reduction at the \emph{Ecosystem library/framework} layer (48 to 23, $-52.1\%$), followed by \emph{Application security logic/policy} (111 to 85, $-23.4\%$). This indicates that the strongest differences occur where mitigations are expressed through framework configuration and application-level decisions.

Other layers are small. \emph{Platform/OS API} decreases from 2 to 1, while \emph{Standard runtime/library} increases from 1 to 2 and should be interpreted cautiously. In this Java Spring Boot setting, runtime misuse is rare and may be more prominent in lower-level ecosystems (e.g., C/C++). No findings are assigned to \emph{Language-core} or \emph{Toolchain/optimization}.

\begin{takeaway}
The MA-CoT shows that pre/post differences concentrate at the framework and application-logic layers.
\end{takeaway}

\begin{table}[t]
\centering
\caption{Weaknesses mapped to the MA-CoT taxonomy.}
\label{tab:layer_attribution}
\begin{tabular}{lrrr}
\toprule
Architectural layer & Pre & Post & Change \\
\midrule
Application security logic/policy & 111 & 85 & $-$23.4\% \\
Ecosystem library/framework & 48 & 23 & $-$52.1\% \\
Platform/OS API & 2 & 1 & $-$50.0\% \\
Standard runtime/library & 1 & 2 & +100.0\% \\
\midrule
Total & 162 & 111 & $-$31.5\% \\
\bottomrule
\end{tabular}
\end{table}

\BfPara{Discussion}
Results indicate that post-training runs are associated with lower validated security burden in LLM-assisted backend development, with strongest evidence at the participant level on the severity-weighted endpoint. The largest descriptive differences occur in authorization and object access, and in authentication, credential policy, and recovery, which are central failure modes in identity-centric systems and emphasized in training. Smaller differences appear in session, request, and browser trust-boundary issues, as well as in abuse-resistance and implementation-level hardening that often depend on framework defaults, rate limiting, or configuration rather than prompt structure. Given the fixed pre-to-post sequence, findings should be interpreted as quasi-experimental associations, not causal proof. The expertise comparison is exploratory. Both groups show reduced burden post-training, but the sample is too small to support an expertise-by-package interaction. Practically, structured developer-side guidance may help express security constraints more effectively, but it complements rather than replaces static analysis, expert review, secure defaults, and deployment hardening.

From a machine learning perspective, these results highlight the importance of interaction-level alignment in generative systems. Rather than modifying models, improvements can be achieved by shaping how users communicate constraints. This suggests that human-centered interventions may play a complementary role alongside model-centric approaches in improving the reliability and security of LLM outputs.

\section{Threats to Validity}\label{sec:threat_to_validity}

\BfPara{Construct Threats}
The primary outcome is the security burden of submitted code, assessed via a consolidated review by the first and second authors using a predefined guide. Manual review is not a perfect proxy for real-world exploitability: some weaknesses may be missed and severity may vary by context. To mitigate this, we required independent review followed by structured adjudication. Because reviewers are authors, residual bias may remain.

\BfPara{Internal Threats}
The fixed pre-to-post sequence may introduce practice or familiarity effects. We mitigate this via counterbalanced task sets, calibrated tasks, and resetting to the same clean snapshot for each run. The training package is standardized in content, delivery, and duration logging. All participants use the same project skeleton, interface, and LLM configuration, reducing extraneous variation. The post-training condition is a bundled intervention (instructional content, MA-CoT model, structured prompting). We cannot isolate the contribution of each component. Results should be interpreted as evidence on the combined package under these conditions. 

We also note that the design does not support causal identification due to the fixed ordering of conditions. Results should therefore be interpreted as quasi-experimental associations rather than definitive causal effects.

\BfPara{External Threats}
The study is conducted in a specific setting: a Java Spring Boot backend with a fixed LLM configuration. This improves control but limits generalizability to other languages, frameworks, domains, models, or environments. Prior work shows that security-related LLM performance varies with contextual factors such as context-window conditions, motivating replication under additional model configurations and prompt settings~\cite{LinD25}. The sample is modest and grouped by broad experience levels. Findings should be interpreted within these constraints.

We note that while the study focuses on a Java Spring Boot backend, the underlying mechanism of structured expression of security constraints during LLM interaction is not tied to a specific language or framework and is expected to generalize conceptually, though empirical validation is required.

\BfPara{Conclusion}
The sample size limits power for subgroup and interaction effects. We emphasize paired participant-level comparisons and treat expertise analyses as exploratory. Validity is strengthened through independent dual review with structured adjudication. Acceptance tests and build checks ensure evaluable submissions only and are not security endpoints.

\section{Conclusion and Future Work}\label{sec:conclusion}
This paper presents a quasi-experimental developer study on post-training effects in LLM-assisted Java backend implementation. Post-training runs show fewer validated weaknesses and lower severity-weighted burden (162 to 111; 432 to 267), with the strongest evidence at the participant level. Gains center on authorization and authentication-related weaknesses, with limited change in session and abuse-resistance controls. Expertise effects remain exploratory. The contribution is practical: structured developer-side training can improve how security constraints are expressed to an LLM without modifying the model. It complements, not replaces, secure defaults, code review, static analysis, and operational hardening. MA-CoT analysis further shows that improvements center on framework and application-logic layers. The key implication is that improving the security of LLM-generated code does not necessarily require better models, but better interactions.

\BfPara{Future Work}
Future work includes analyzing interaction logs for effective prompting patterns~\cite{TonyFMDS25}, larger replications for expertise and order effects, and extensions to other languages, frameworks, and models~\cite{KharmaCAD25,CroftXZBT22}. Training should also target persistent weaknesses, including operational hardening, rate limiting, secure configuration, and log hygiene.


\appendices
\section{Participant-level severity profiles}\label{app:participant_profiles}
Table~\ref{tab:participant_severity_profiles} reports raw severity composition for each participant in both conditions. Its purpose is transparency; formal analysis is based on paired participant-level change scores.

\begin{table}[!htbp]
\centering
\caption{Participant-level severity profile before and after training. C=critical, H=high, M=medium, L=low, and T=total.}
\label{tab:participant_severity_profiles}
\scriptsize
\begin{tabular}{llrrrrrrrrrr}
\toprule
 &  & \multicolumn{5}{c}{Pre-training} & \multicolumn{5}{c}{Post-training} \\
\cmidrule(lr){3-7} \cmidrule(lr){8-12}
ID & Stratum & C & H & M & L & T & C & H & M & L & T \\
\midrule
P01 & Senior & 1 & 6 & 0 & 1 & 8 & 0 & 4 & 0 & 1 & 5 \\
P02 & Senior & 3 & 8 & 5 & 1 & 17 & 0 & 2 & 0 & 0 & 2 \\
P03 & Senior & 1 & 7 & 4 & 3 & 15 & 1 & 9 & 6 & 2 & 18 \\
P04 & Junior & 5 & 8 & 2 & 1 & 16 & 0 & 3 & 4 & 1 & 8 \\
P05 & Junior & 1 & 10 & 4 & 1 & 16 & 2 & 7 & 0 & 3 & 12 \\
P06 & Junior & 2 & 8 & 3 & 1 & 14 & 1 & 2 & 3 & 1 & 7 \\
P07 & Junior & 3 & 4 & 6 & 1 & 14 & 1 & 6 & 7 & 0 & 14 \\
P08 & Senior & 0 & 7 & 1 & 2 & 10 & 0 & 6 & 3 & 1 & 10 \\
P09 & Senior & 3 & 8 & 2 & 4 & 17 & 0 & 7 & 5 & 1 & 13 \\
P10 & Senior & 3 & 4 & 2 & 0 & 9 & 0 & 6 & 2 & 6 & 14 \\
P11 & Junior & 0 & 3 & 7 & 4 & 14 & 0 & 0 & 3 & 3 & 6 \\
P12 & Junior & 2 & 7 & 2 & 1 & 12 & 0 & 2 & 0 & 0 & 2 \\
\midrule
Total & -- & 24 & 80 & 38 & 20 & 162 & 5 & 54 & 33 & 19 & 111 \\
\bottomrule
\end{tabular}\vspace{-3mm}
\end{table}


\subsection{CWE-to-Weakness-Family Mapping}
\label{app:cwe_family_mapping}

To support the descriptive domain-level analysis in Table~\ref{tab:cwe_family_mapping}, each CWE was assigned to a single researcher-defined weakness family. The mapping reflects recurring security concerns in identity-centric backends while reducing fragmentation across low-frequency CWEs. Because some assignments are interpretive, it is used only for descriptive summarization and not for participant-level inference.

\begin{table}[H]
\centering
\caption{CWE-to-weakness-family mapping used for the domain-level analysis.}
\label{tab:cwe_family_mapping}
\scalebox{0.64}{
\begin{tabular}{p{5.2cm} p{7.8cm}}
\toprule
Weakness family & CWE identifiers \\
\midrule
Authorization \& object access & CWE-269, CWE-639, CWE-732, CWE-807, CWE-862 \\
Authentication, credential policy \& recovery & CWE-258, CWE-259, CWE-287, CWE-306, CWE-521, CWE-522, CWE-523, CWE-640, CWE-798, CWE-916 \\
Session, request \& browser trust boundary & CWE-352, CWE-384, CWE-613, CWE-1021 \\
Sensitive data, secrets \& cryptography & CWE-201, CWE-312, CWE-319, CWE-321, CWE-330, CWE-338 \\
Logging, errors \& information exposure & CWE-117, CWE-203, CWE-204, CWE-209, CWE-532, CWE-778 \\
Configuration \& debug surface & CWE-489 \\
Input validation \& unsafe integration & CWE-20, CWE-611, CWE-90, CWE-918 \\
Availability \& abuse resistance & CWE-307, CWE-400 \\
File handling & CWE-434 \\
\bottomrule
\end{tabular}
}
\end{table}

\end{document}